\documentclass{llncs}
\usepackage[margin=1.0in]{geometry}

\usepackage{times}
\SetMathAlphabet{\mathtt}{normal}{OT1}{pcr}{n}{n}
\SetMathAlphabet{\mathtt}{bold}{OT1}{pcr}{bx}{n}

\usepackage{amsmath}
\usepackage{amssymb} 

\newcommand{\CUT}[1]{}

\newcommand{\secref}[1]{Section~\ref{#1}}
\newcommand{\tblref}[1]{Table~\ref{#1}}
\newcommand{\figref}[1]{Figure~\ref{#1}}

\newcommand{\eg}{{\em e.g.}}

\newcommand{\ie}{{\em i.e.}}

\newcommand{\naive}{na\"{\i}ve}

\newcommand{\etal}{{\em et al.\/}}
\newcommand{\role}{r\^{o}le}

\newcommand{\SET}[1]{\ensuremath{\{#1\}}}

\newcount\timeHH
\newcount\timeMM
\timeHH=\time
\divide\timeHH by 60
\timeMM=\time
\multiply\count255 by -60 \advance\timeMM by \count255
\newcommand{\timestamp}{%
  \today{} ---
  \ifnum\timeHH<10 0\fi\number\timeHH\,:\,\ifnum\timeMM<10 0\fi\number\timeMM}

\usepackage{color}
\definecolor{Red}{rgb}{0.9,0.0,0.0}  
\definecolor{Green}{rgb}{0.0,0.4,0.0}
\definecolor{Blue}{rgb}{0.0,0.0,0.9}
\definecolor{DarkBlue}{rgb}{0.0,0.0,0.75}
\definecolor{Midnight}{rgb}{0.0,0.0,0.5}
\definecolor{Purple}{rgb}{0.5,0.0,0.4}
\definecolor{Black}{rgb}{0.0,0.0,0.0}
\definecolor{Yellow}{rgb}{1.0,1.0, 0.25}
\definecolor{Cyan}{rgb}{0.25,1.0, 1.0}

\newcommand{\cdColor}{Black}
\newcommand{\kwColor}{DarkBlue}
\newcommand{\comColor}{Red}

\usepackage{listings}
\lstdefinelanguage{Diderot}{%
  otherkeywords={|,\#},
  morekeywords={%
    bool,%
    die,%
    else,%
    false,field,foreach,%
    identity,if,image,in,inf,initially,input,int,%
    kernel,%
    nan,new,%
    output,%
    real,%
    sphere,stabilize,strand,string,%
    tensor,true,%
    update,%
    vec2,vec3,vec4,%
    zeros},
  morekeywords=[2]{D},
  sensitive,%
  morecomment=[s]{/*}{*/},%
  morecomment=[l]//,
  morestring=[b]"}%

\usepackage{amsmath}
\usepackage{mathtools}

\newcommand{\name}{EIN}

\newcommand{\fontR}{\mathbb{R}}

\newcommand{\optshift}{\textit{Shift}}

\newcommand{\optsplit}{\textit{Split}}
\newcommand{\optslice}{\textit{Slice}}

\newcommand{\fontrelated}[1]{\textit{#1}}

\newcommand{\EinExp}[2]{\ensuremath{{\left\langle{#1}\right\rangle}_{#2}}}
\newcommand{\EinOp}[3]{\ensuremath{\lambda{}\,#1\EinExp{#2}{#3}}}
\newcommand{\EinApp}[4]{\ensuremath{\EinOp{#1}{#2}{#3}(#4)}}

\newcommand{\lift}[1]{\textbf{lift}({#1})}

\newcommand{\einpart}[1]{\frac{\partial}{\partial x_{#1}}\diamond}

\newcommand{\mlstinline}[1]{\text{\lstinline!#1!}}

\title{Compiling Diderot: From Tensor Calculus to C}

\author{
  Charisee Chiw \and
  Gordon L. Kindlmann \and
  John Reppy}

\begin{document}
\maketitle

%

\begin{abstract}
  Diderot is a parallel domain-specific language for
  analysis and visualization of multidimensional scientific images,
  such as those produced by CT and MRI scanners~\cite{diderot:pldi12,diderot:vis15}.
  In particular, it supports algorithms where tensor fields (\ie{}, functions from
  3D points to tensor values) are used to represent the underlying physical objects that
  were scanned by the imaging device.
  Diderot supports higher-order programming where tensor fields are first-class
  values and where differential operators and lifted linear-algebra operators
  can be used to express mathematical reasoning directly in the language.
  While such lifted field operations are central to the definition and
  computation of many scientific visualization algorithms, to date they
  have required extensive manual derivations and laborious
  implementation.

  The challenge for the Diderot compiler is to effectively translate the
  high-level mathematical concepts that are expressible in the surface
  language to a low-level and efficient implementation in C.
  This paper describes our approach to this challenge, which is based
  around the careful design of an intermediate representation (IR), called EIN,
  and a number of compiler transformations that lower the program from
  tensor calculus to C while avoiding combinatorial explosion in the
  size of the IR.
  We describe the challenges in compiling a language like Diderot, the
  design of EIN, and the transformation used by the compiler.
  We also present an evaluation of EIN with respect to both compiler
  efficiency and quality of generated code.
\end{abstract}%

\section{Introduction}

Diderot is a domain-specific language for writing image-analysis algorithms
that are defined using the concepts of differential tensor calculus and linear
algebra~\cite{diderot:pldi12,diderot:vis15}.
The design of Diderot couples a simple portable parallelism model with a very-high-level
mathematical programming notation, which allows a
user to directly transfer her mathematical reasoning from whiteboard to code.
For example, a CT scan can be viewed as a 3D scalar tensor field $F$ (\ie{}, a continuous
map from $\Re^3$ to scalar values), where the expression $F(x)$ 
represents the opacity to X-rays of the scanned object at point $x$.
Using the concepts of tensor calculus, we can compute geometric properties of the scanned object.
For our example, if the point $x$ lies on an isosurface in $F$, such as
the boundary between hard and soft materials, the surface-normal vector at $x$
can be computed using the expression
$-\nabla{}F(x)/|\nabla{}F(x)|$.

As in mathematics (and in functional programming), Diderot lifts operations on tensors
(\eg{}, inner product) to work on tensor fields.
Lifted operations allow programmers a more concise expression of their algorithmic ideas and
allows them to avoid the tedious and error-prone by-hand lowering of higher-order operations
to first-order code~\cite{chiw:phd-thesis}.
In these ways, Diderot provides a very-high-level mathematical programming model for computing
with 2D and 3D image data.

The challenge of implementing Diderot is bridging the wide semantic gap from its
very-high-level programming notation to the low-level code produced by the compiler.
This paper describes an effective approach to this challenge that is based on an intermediate
representation called \name{} and a complementary set of optimizations and transformations.
Many optimizing compilers rely on a normalized representation (\eg{},
SSA~\cite{efficiently-computing-ssa},
ANF~\cite{essence-of-comp-w-conts}, or CPS~\cite{appel:cps-book}) where operations are applied to
atomic values and intermediate results are given names, while other compilers have used expression
trees to represent computation~\cite{lcc-book,ghc-inliner}.
In the design of \name{}, we take a hybrid approach, where we embed an extensible family of
operators that are defined by indexed expression trees inside an SSA-based IR.
As we show, this hybrid approach allows a concise representation of the nested loops that
arise in tensor computations, it supports the transformations necessary for implementing
Diderot's higher-order features, and it enables easier extension of the language, thus providing
a solid foundation for future growth of the language.

The remainder of the paper is organized as follows.
In the next section, we provide necessary background material, including a description of the
computational core of Diderot, the mathematical foundations of this core, and an outline of the
Diderot compiler's architecture.
We then describe the motivation for the work presented in the paper in \secref{sec:motivation}.
Sections~\ref{sec:design} and~\ref{sec:implementation} are the cover the main results of the paper:
the design of our IR and its implementation in the compiler.
In \secref{sec:benchmarks}, we present an empirical evaluation of the compiler and the various
choices of optimization.
The main benefit of this work is that it supports a much more expressive programming model;
in \secref{sec:applications} we describe two algorithmic techniques that are enabled by the
\name{} IR.
Related work is described in \secref{sec:related} followed by a conclusion.


\section{Diderot}
\label{sec:background}
In this section, we given an overview of the Diderot language and explain the
underlying techniques for representing continuous tensor fields that are
reconstructed from multidimensional image data.

\subsection{The Diderot Language}

For this paper, we are primarily concerned with the computational core of Diderot; more
complete descriptions of the language design can be found in previous
papers~\cite{diderot:pldi12,diderot:vis15}, and tutorial examples can be
found on the Diderot example repository on
GitHub (\url{https://github.com/Diderot-Language/examples}).
The computational core is mostly organized around two families of types:
\begin{description}
  \item[tensors] are concrete values that include scalars (0th-order),
    vectors (1st-order), matrices (2nd-order), and higher-order tensors.
    The Diderot type ``\lstinline[mathescape=true]!tensor[$d_1,\,\ldots,\,d_n$]!''
    is the type of $n$th-order tensors in $\fontR{}^{d_1} \times \cdots \times \fontR{}^{d_n}$,
    where $n \geq 0$.
    We refer to $d_1,\,\ldots,\,d_n$ as the \emph{shape} of the tensor and will use
    $\sigma$ to denote shapes when the individual dimensions are not important.\footnote{
      Note that we exclusively use the orthonormal elementary basis for representing
      tensors, which means that covariant and contra-variant indices can be treated equally.
    }
    Also, we require that the dimensions $d_i$ satisfy $d_i > 1$ to ensure that shapes are canonical.
  \item[tensor fields] are continuous functions from $\fontR{}^{d}$, where $1 \leq d \leq 3$,
    to tensors.
    The Diderot type ``\lstinline[mathescape=true]!field#$k$($d$)[$\sigma$]!'' is the type of
    tensor fields that map values in $\fontR{}^{d}$ to tensors of type
    \lstinline[mathescape=true]!tensor[$\sigma$]!.
    The value $k > 1$ specifies the \emph{continuity} of the field; \ie{}, how many times
    it can be differentiated.
\end{description}

Diderot supports standard linear algebra operations on tensors, such as addition,
subtraction, dot products, as well as other tensor operations such as
double dot (colon) products, outer products, and trace.
Diderot's expression syntax is designed to look similar to mathematical notation,
while still retaining the flavor of a programming notation.
It uses Unicode math symbols, instead of ASCII identifiers, for many operators.
For example, one writes the expression
\begin{center}
  \lstinline[mathescape=true]!(u $\otimes$ v) / |u $\otimes$ v|!
\end{center}%
for the normalized outer product of two tensors.

In addition to tensors and fields, Diderot supports \emph{images}, which, like fields,
have a dimension and a shape.
Programs do not compute directly with images, instead we convolve images with \emph{kernels}
to form an initial tensor field.
The resulting field gets its dimension and shape from the image and its continuity from
the kernel.
Higher-order operators can then be applied to these values to define derived tensor fields.
The operations on tensor fields include various differentiation operators,
and lifted linear-algebra operations.
In addition, a field can be applied to a point to yield a tensor value (we call this
operation \emph{probing} the field at the point).

\subsection{Implementing Tensor Fields}
\label{back:reconstruction}

We classify operations on fields as either \emph{declarative}, \ie{}, operations that
define field values, or \emph{computational}, \ie{}, operations that query a field to
extract a concrete value.\footnote{
  There are two computational operations on fields: testing if a point is in the domain
  of a field and probing.
  We focus on the latter in this paper.
}
The compilation process uses the field definitions produced by declarative operations to
generate code that implements the computational operations on those fields.
In this section, we give an informal description of the mathematics of implementing
probes.

\begin{figure}[t]
  \begin{center}
    \includegraphics[width=4.2in]{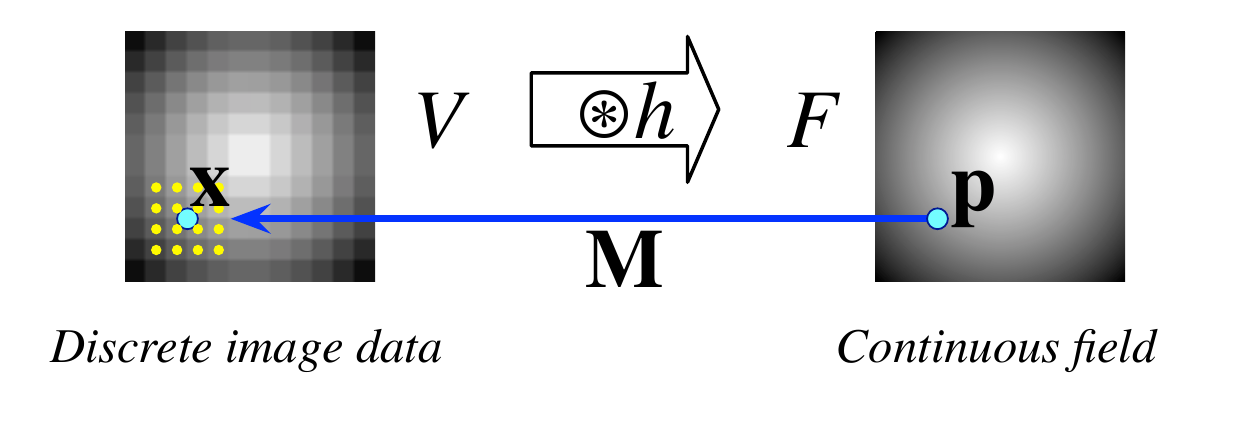}
  \end{center}%
  \caption{Probing a 2D field $F = h \circledast{} V$ at $\textbf{p}$,
    where $\textbf{M}$ is the mapping from
    world coordinates to image-space coordinates}
  \label{fig:probe}
\end{figure}

To start, consider a scalar 2D field $F$ that is defined as the convolution $H \circledast V$
of an image $V$ with a reconstruction kernel $H$, where $H$ is a separable kernel
function over multiple arguments (\ie{}, $H(x,y) = h(x)h(y)$).
Probing the field $F$ at a point $\mathbf{p}$ involves mapping $\mathbf{p}$ to
a region of $V$ and then computing a weighted sum of the voxel values in the region,
where the kernel $H$ provides the weights.
Let $\mathbf{M}$ be the mapping from world coordinates to image space.
Then $\mathbf{x} = \mathbf{M}\mathbf{p}$ is the point in image space and
$\textbf{n} = \lfloor \textbf{x} \rfloor$
defines the voxel indices for $\mathbf{p}$ in the image data.
Then we can compute the probe of $F$ at $\mathbf{p}$ as follows:
\begin{displaymath}
  F(\mathbf{p}) = (H \circledast V)(\mathbf{p}) = \sum^{s}_{i=1-s} \sum^{s}_{j=1-s}
  \left(
    h(\textbf{f}_0-i) h(\textbf{f}_1-j) V[\textbf{n}_0+i,\textbf{n}_1+j]
  \right)
\end{displaymath}%
where the \emph{support} of the kernel $H$ is $2s$ and $\textbf{f} = \textbf{x} - \textbf{n}$.
\figref{fig:probe} illustrates this computation for a kernel with support $4$.

Handling differentiation of fields requires pushing the differentiation operators down
to the convolutions, where they can be applied to the kernel functions.
Symbolically, this transformation is
\begin{displaymath}
  \nabla{}F(\mathbf{p}) \Rightarrow \nabla{}(H \circledast V)(\mathbf{p})
    \Rightarrow ((\nabla{} H) \circledast V)(\mathbf{p})
\end{displaymath}%
Because we use separable kernels, their differentiation is straightforward:
\begin{displaymath}
  \nabla H(x,y)=
    \left[\begin{array}{c}
      \frac{\partial}{\partial_x} \\[0.5em]
      \frac{\partial}{\partial_y}
    \end{array}\right] H(x,y) =
    \left[\begin{array}{c}
      \frac{\partial}{\partial_x} H(x,y) \\[0.5em]
      \frac{\partial}{\partial_y} H(x,y)
    \end{array}\right] =
    \left[\begin{array}{c}
      \frac{\partial}{\partial_x} (h(x) h(y)) \\[0.5em]
      \frac{\partial}{\partial_y} (h(x) h(y))
    \end{array}\right] =
    \left[\begin{array}{c}
      h'(x) h(y) \\[0.5em]
      h(x) h'(y)
    \end{array}\right]
\end{displaymath}%
where $h'$ is the first derivative of $h$.

Thus, the process of implementing a probe operation requires expanding the definition
of the field and then \emph{normalizing} the resulting expression
so that code for the appropriate reconstruction can be generated.
As we saw above, normalization involves pushing differential operators down to the
initial-field definitions.
It also involves distributing differential operators and
pushing probes down below lifted operators.
In effect, part of normalization involves lowering higher-order operators to their first-order
equivalents.
For example, the probe of the gradient sum of two fields is normalized to the sum of probes
of gradients:
\begin{displaymath}
  \nabla(F + G)(x) \Rightarrow (\nabla{}F + \nabla{}G)(x)
    \Rightarrow \nabla{}F(x) + \nabla{}G(x)
\end{displaymath}%
which can then be handled as above.
The normalization process is a necessary part of the translation of Diderot into
executable code.

\subsection{Direct vs.\ index-sensitive operators}
We find it is useful to make a distinction between what
we call \emph{direct-style} operators and those that we describe as \emph{index-sensitive}.
This distinction is important for comparing the expressiveness of
different implementation approaches, but is not visible in the surface language.
Roughly speaking, the mathematical definition of a direct-style operations does not depend
on the choice of a coordinate system (\eg{}, the dot product is mathematically defined
in terms of the cosine of the angle between vectors), whereas the definition of
index-sensitive operations relies on some choice of coordinate system and the indices
into the individual coordinates of quantities expressed in that coordinate system.
From the point of view of the implementation, index-sensitive operations require
explicit manipulation of indices.
A common example of an index-sensitive operation is the curl, which is discussed in
\secref{sec:design} (see \figref{fig:ein-examples}).

\subsection{Compiler architecture}
\label{sec:compiler}

This paper is, in part, a tale of two compilers:
the first Diderot compiler~\cite{diderot:pldi12}, which suffered from various
limitations, and the current compiler, which addresses those limitations.
Both of these compilers have the same basic architecture, which we describe here.

The Diderot compiler is organized into three main phases: the front-end,
optimization and lowering, and code generation.
The front-end handles parsing, type checking, and some preliminary optimizations,
and produces a simplified monomorphic representation, called Simple AST, where temporaries are
introduced for intermediate values and operators are applied only to variables.
The middle phase lowers the surface-language operations (\eg{}, field probes) to
low-level code (\eg{}, memory and arithmetic operations to reconstruct field values).
It consists of three stages, where each stage manipulates a different intermediate
representation.
These IRs share a common Static Single Assignment (SSA) form~\cite{efficiently-computing-ssa},
but different in their types and operations.
\begin{description}
  \item[HighIR] is essentially a desugared version of the source language with
    source-level types and operations.
  \item[MidIR]
    supports vectors, transforms between coordinate spaces, loading image data,
    and kernel evaluations.
    At this stage, fields and probes have been compiled away into first-order code.
  \item[LowIR]
    supports basic operations on vectors, scalars, and memory objects.
\end{description}%
The translations between these representations replaces higher-level operations with their
equivalent lower-level operations.
For example, field probes in the HighIR (after normalization) are expanded into
the convolution of image samples and kernel evaluations in the MidIR.
We apply a value-numbering pass to eliminate redundant computation on each of the IRs.
For our domain, redundant computations naturally arise in many places; for example,
the Hessian produces symmetric matrices and reconstruction of field
overlaps with the reconstruction of its gradient.
The final phase takes the LowIR representation and converts it to expression trees
and statements, which are emitted as vectorized C code and then compiled to either
a standalone executable or a library that can be embedded in a larger application.

In this paper, we are primarily interested in the representation of operations in
the optimization and lowering phase.


%
\section{Motivation}
\label{sec:motivation}

In programming, as in mathematics, there is a great expressiveness benefit in being
able to \emph{lift} simple operations to work on higher-order values.
In Diderot's domain of image analysis and visualization algorithms, we see this benefit
in the lifting of linear-algebra operations (\eg{}, vector-norm or dot product) to work
on tensor fields.

For example, accurate and robust detection of surfaces (or edges) in images is
an important part of volume-rendering applications.
A classical principle of edge detection proposed by Canny is that edges
are where the image gradient magnitude is maximized with respect to motion
along the gradient direction~\cite{canny1986}.
In a scalar tensor field $F$, the locations where $|\nabla F(x)|$
is maximized with respect to motion along $\nabla F(x)/|\nabla F(x)|$ are where
$\nabla|\nabla F(x)|$ (the gradient of quantity being maximized)
is orthogonal to $\nabla F(x)/|\nabla F(x)|$.
In other words, the set of points where the field
\begin{displaymath}
  E = -\nabla |\nabla{} F| \mathbin{\bullet{}} \frac{\nabla{} F}{|\nabla{} F|}
\end{displaymath}%
is zero (\ie{}, $E(x) = 0$).
Notice that the definition of $E$ is higher-order; \ie{} it is purely in terms of fields
and higher-order operations on fields.
We believe that this kind of higher-order notation is the most natural way for an
image-analysis expert to reason about and develop image-analysis algorithms.

The first implementation of Diderot~\cite{diderot:pldi12} had some significant
limitations in how well it supported higher-order programming with tensor fields.
Specifically, it could not fully support index-sensitive operations, such as
the curl on fields ($\nabla{}{\times}$) and it only lifted a very limited
set of linear-algebra operations to work on fields (just scalar multiplication,
field addition, and field subtraction), which meant that the style of higher-order
programming described above was not possible.
Instead, it was necessary for a programmer to derive by hand the first-order
implementation of the higher-order reasoning.
Such a process is time-consuming, error-prone, and can result in a significant
increase in the size of the code~\cite{chiw:phd-thesis}.

The limitations on higher-order operations were owed to the representation of operations
in the previous compiler's IR.
As described in \secref{sec:compiler}, the compiler uses an SSA representation for
the optimization and lowering phase.
In the previous compiler, we defined a fixed set of \emph{direct-style} operators,
such as \texttt{DotProduct}, \texttt{FieldAdd}, \texttt{FieldGrad}, for the HighIR,
which corresponded to the surface-language operators.
This representation allowed a straightforward implementation of normalization and other
transformations using term rewriting~\cite{diderot:pldi12}.
There was a problem, however, in that adding a single new field-level operator would require
adding additional rewrite rules for the possible combinations of that operator and the other
field operations, as well as operator-specific lowering code and down-stream support for
the operator.
Essentially, every lifted operator would be a special case in the IR.

To address these limitations we developed a new intermediate representation, called \name{},
for the Diderot compiler.
This representation allows a wide range of higher-order computations, including
expressions like
\begin{lstlisting}[mathescape=true]
    field#2(3)[] E = -$\nabla$(|$\nabla$F|) $\bullet$ $\nabla$F/|$\nabla$F|;
\end{lstlisting}
for the edge detection example.
The \name{} representation has enabled a new level of flexibility and expressiveness
in the Diderot language.
This expressiveness makes developing sophisticated new algorithms easier, faster, and more intuitive.
Diderot programmers are able to express their ideas in a high-level mathematical notation and then
rely on the compiler to handle the necessary derivations to realize an implementation.
We present two examples of the benefit of this expressiveness in \secref{sec:applications}.

%
\section{The design of EIN}
\label{sec:design}

The desire to support higher-order programming with tensor fields required a
redesign of the exiting IR.
The resulting design needed to satisfy several requirements:
\begin{enumerate}
  \item it must support the normalization transformations that are needed
    to turn probes of arbitrary fields into realizable computations;
  \item it must provide a mechanism to express index-sensitive operations
    on fields (\eg{}, curl); and
  \item it must be easily extensible to new operations on both tensors and fields.
\end{enumerate}%

The existing direct-style operators in the previous compiler only satisfied the
first of these requirements.
While it was possible to add operations like curl as IR operators, they would have
to be treated as black boxes by the compiler and
their use prevents further differentiation of the result.
Lastly, adding new operations on field required adding new normalization rewrites
for all of the existing field operations, as well as new lowering and code generation
code.

An alternative approach that we considered was immediately expanding the tensor structure into
loops or unrolled straight-line code in the SSA.
This approach has the major drawback that the normalization transformations become
more difficult; the cost of normalization is higher, since each component of a field-valued
expression is normalized separately, and the size of the IR might explode.\footnote{
  Our subsequent experience with implementing \name{} demonstrated that this last concern
  was a significant problem in practice.
}

Although there were drawbacks to the previous compiler's IR; there were positive aspects as well.
The SSA-representation provided a good basis for standard compiler analyses and optimizations,
such as value numbering.
Also direct-style operators abstract away from the implicit iterative structure required to
implement them, which helped keep the size of the IR manageable during compilation.
Only at the last stage, after much redundancy had been eliminated, did we expand out the
iterative code into vectorized operations.
In designing a new IR, we wanted to preserve these positive features as much as possible,
while adding new flexibility to meet the other design goals.
Our solution is to keep the existing SSA representation, but to replace the tensor and
tensor-field operators with a family of operators defined by expression trees.
We add a new form of SSA operator, called an \name{} operator,\footnote{
  The name \name{} was chosen because the syntax for these operators was
  originally inspired by \emph{Einstein Index Notation},
  which is a concise written notation for tensor calculus
  invented by Albert Einstein~\cite{Einstein:general-theory}.
} which has the form
$\EinOp{\mathit{xs}}{e}{\sigma}$, where
\begin{itemize}
  \item
    $\mathit{xs}\in\textsc{Var}^{*}$ are the parameters of the operator;
  \item
    $e$ is an \name{} expression that is the body of the operator; and
  \item
    $\sigma \in (\textsc{IndexVar} \times \mathbb{N})^{*}$ defines the
    \emph{index space} of the operator, where the first component of the pairs
    are the index variables and the second components are the upper bounds (the
    lower bounds are $1$ by definition).
    If $\sigma = i_1 \leq d_1,\, \ldots,\, i_k \leq d_n$, then the shape of the
    tensor (or tensor field) defined by the operator is $d_1, \ldots, d_n$.
\end{itemize}%
One can think of the index space as defining an $n$-deep loop nest over the index variables
where the expression $e$ is the loop body that defines scalar components of the result.
The scope of the parameters and index variables is $e$.
At the SSA level, these operators are applied to arguments on the right-hand-side of
assignments; \eg{}, $t = \EinApp{\mathit{xs}}{e}{\sigma}{\mathit{ys}}$.
By keeping the operator definition separate from the arguments, optimizations at
the SSA level can analyze and transform these assignments without having to
examine the body $e$ of the \name{} operator.

\begin{figure}[t]
  \begin{displaymath}
    \begin{array}{rclr}
      \textbf{E} & ::= & \EinOp{\bar{x}}{e}{\sigma} & \text{\name{} Operator} \\[0.5em]
       e & ::=  & T_{\alpha} & \text{Indexed tensor parameter}\\
         & \mid & F_{\alpha} & \text{Indexed field parameter}\\
         & \mid & \delta_{ij} & \text{Kronecker delta}\\
         & \mid & \mathcal{E}_{ij} \;\mid\; \mathcal{E}_{ijk} & \text{Levi-Civita tensor}\\
         & \mid & \sin(e) \;\mid\; \arcsin(e) \;\mid\, \ldots \,\mid\; \arctan(e)
                & \text{Trig.\ functions} \\
         & \mid & \sqrt{e} \;\mid\; -e \;\mid\; \exp(e) \;\mid\; e^n & \text{Unary operators} \\
         & \mid &  e+e \;\mid\; e-e \;\mid\; e*e \;\mid\; \frac{e}{e} & \text{Binary operators} \\
         & \mid & \sum\limits_{n\leq{}i\leq{}n'} e & \text{Summation}\\  [0.25em]
         & \mid &  e_1 @ e_2 & \text{Probe of a field $e_1$ at position $e_2$} \\
         & \mid &{\lift{e}}& \text{Lift tensor $e$ to a field}\\
         & \mid & V_{\alpha} \circledast H^\beta & \text{Convolution}\\
         & \mid & \einpart{i}  e  & \text{Partial derivative of $e$}\\[0.5em]
      \sigma & \in &(\textsc{IndexVar} \times \mathbb{N})^{*}
             & \text{Index map} \\[0.5em]
      i,j,k & \in & \textsc{IndexVar} & \text{Variable index} \\[0.5em]
      n & \in & \mathbb{N} & \text{Constant index} \\[0.5em]
      \mu & \in & \textsc{Index} = \textsc{IndexVar} \cup \textsc{ConstVar} & \text{Single index} \\[0.5em]
      \alpha,\beta,\gamma & \in & \textsc{Index}^{*} & \text{Sequence of indices} \\
    \end{array}
  \end{displaymath}%
  \caption{The syntax of \name{} operators and \name{} expressions in HighIR}
  \label{fig:ein-syntax}
  \vspace{-2em}
\end{figure}%

The syntax of \name{} operators and expressions is given in \figref{fig:ein-syntax}.
An expression can be an indexed tensor ($T_{\alpha}$) or tensor-field ($F_{\alpha}$)
parameter ($\alpha$ defines which component of the tensor is referenced).
The Kronecker delta and Levi-Civita tensors are used to are used to permute and
cancel components based on their indices.
\name{} expressions also include the trigonometric functions and standard
arithmetic operators.
The summation expression introduces an additional index variable and is the only binding
form in the \name{} expressions.
We use the syntax $e_1 @ e_2$ to represent the probe of a field defined by $e_1$ by the
value $e_2$.
The expression $\lift{e}$ expression lifts a tensor value to be a constant field.
Initial fields are defined by convolution $V_{\alpha} \circledast H^{\beta}$
of an image $V$ with a kernel $H$, where $\alpha$ indexes the shape of voxels in
the image (\eg{}, if $V$ is a vector image, then $\alpha$ will be a single index
ranging over the vector indices) and $\beta$ specifies the differentiation of the
kernel.
Lastly, the application of the partial derivative with respect to the $i$th axis
of an expression is represented as $\einpart{i} e$.
\name{} expressions are also used in the MidIR, but at that point fields have been
rewritten to tensor computations and so the syntactic forms for field expressions are not used.

Index maps ($\sigma$) and indices play an important \role{} in the syntax and semantics
of \name{}.
Indices can either be variables (denoted by $i$, $j$, and $k$), or constants ($n\in \mathbb{N}$).
We use $\alpha$ and $\beta$ to denote sequences of zero or more indices of either type.
When describing transformations it will be useful to connect the variables bound in an
index map with the indices of a tensor or field variable.
We use the notation $\hat{\alpha}$ to denote an index map, where $\alpha$ is the corresponding
sequence of index variables.

The \name{} representation is a hybrid design that embeds expression trees into a normalized
SSA representation.
This design preserves the useful properties of the SSA representation, while providing
flexibility in the specification of tensor and tensor-field operations.
The key property of this design is that it allows reference to indices in the
body of $e$ (thus supporting index-sensitive operators), while also
providing a compact representation of the nested iteration that is implicit in
the definition of tensors and tensor fields.

To give some intuition for \name{} operators we present a number of examples
of common operators in \figref{fig:ein-examples}.
\begin{figure}[t]
\begin{displaymath}
  \begin{array}{rcll}
    \mathbf{u} \bullet{} \mathbf{v} & \equiv & \EinApp{U, V}{U_i * V_i}{i\leq{}n}{%
      \mathbf{u}, \mathbf{v}} &
      \text{\textit{Inner (dot) product of $n$-vectors}} \\[0.25em]
    \mathbf{u} \otimes{} \mathbf{v} & \equiv & \EinApp{U, V}{U_i * V_j}{i\leq{}n, j \leq{}m}{%
      \mathbf{u}, \mathbf{v}} &
      \text{\textit{Outer product of vectors}} \\[0.25em]
    \mathbf{u} \times{} \mathbf{v} &
      \equiv & \EinApp{U, V}{%
        \sum\limits_{j\leq{}3}\sum\limits_{k\leq{}3} \mathcal{E}_{ijk} * U_{j} * V_{k}}{i\leq{}3}{%
      \mathbf{u}, \mathbf{v}}\quad &
      \text{\textit{Cross product of vectors}} \\[0.25em]
    \mathrm{trace}(\mathbf{M}) & \equiv &
      \EinApp{T}{\sum\limits_{i\leq{}n} T_{i,i}}{}{\mathbf{M}} &
      \text{\textit{Trace of an $n\times{}n$ matrix}} \\[0.25em]
    \mathbf{M}^{T} & \equiv & \EinApp{T}{T_{j,i}}{i\leq{}n, j \leq{} m}{\mathbf{M}} &
      \text{\textit{Matrix transpose}} \\[0.25em]
    \nabla{\times{}} \mathbf{f} & \equiv & \EinApp{F}{%
        \sum\limits_{j\leq{}3}\sum\limits_{k\leq{}3}
          \mathcal{E}_{ijk} * \einpart{j} F_{k}}{i\leq{}3}{%
      \mathbf{f}} &
      \text{\textit{Curl of a vector field}}
  \end{array}%
\end{displaymath}%
\caption{Examples of common operators encoded as \name{} operator applications}
  \label{fig:ein-examples}
\end{figure}%
The last of these examples, is the curl of a vector field;, which is an example of
an index-sensitive operation on fields that was not handled fully in the previous compiler's IR
(satisfying the second of our design goals).
In the next section, we describe the compiler transformations that we apply to \name{}
and show that it satisfies the other two design goals.

%
\section{Implementation}
\label{sec:implementation}

With the design of the \name{} IR we had to also develop new techniques
for optimization and lowering, which we describe in this section.
Switching to \name{} in the compiler enabled a new class of program that were
not supported by the previous compiler.
Unfortunately, it turned out that some of these programs caused an explosion
in the size of the \name{} IR is our first implementation of it.
We describe the techniques for managing the size of the IR during compilation
that de developed.

\subsection{Translation}
\label{sec:translation}

The Diderot compiler generates HighIR, including the \name{} operators,
from the monomorphic Simple AST representation produced by the
front end.
The translation from the Simple AST to HighIR is aided by the definition
of meta-level generic \name{} operators that are parameterized over
types, shapes, and dimensions.\footnote{
  These generic operators are written directly in SML, which is the implementation
  language of the compiler.
}
In the Simple AST, surface-language operations have been mapped to
monomorphic instances of shape-polymorphic operations.
We then can convert Simple AST operations to \name{} operators by
applying the corresponding generic operator to the type/shape/dimension
parameters of the AST operator.
This approach allows many similar operations to be defined by a single
generic operator.
For example, vector-vector, vector-matrix, matrix-vector, and matrix-matrix
multiplication are all instances of the same generic inner-product operator.
This mechanism also makes it fairly easy to extend the language with new
operators.

\subsection{Fusion}
\label{sec:fusion}

The result of the translation from Simple AST to High IR is a program
where every surface operation is represented by an SSA assignment.
To enable transformations of the \name{} expressions, we must first
fuse operations to produce larger expressions that are amenable to
useful transformations.
The basic idea is that if we have two SSA assignments involving
\name{} operators
\begin{displaymath}
  \begin{array}{l}
    t_1 = \EinApp{\mathit{xs}}{e_1}{\sigma_1}{\mathit{ys}} \\
    \cdots \\
    t_2 = \EinApp{\ldots,\, t, \,\ldots}{e_2}{\sigma_2}{\ldots,\, t_1, \,\ldots}
  \end{array}%
\end{displaymath}%
then we want to substitute $e_1$ for $t_1$ in $e_2$, while adding the $\mathit{xs}$
to the parameters of the second operator and the $\mathit{ys}$ to its arguments.

The fusion transformation has the potential to explode the size of the program.
In our experience, this problem does not occur for HighIR in real programs, but
code duplication does cause problems down stream in Mid and Low IR, as we discuss
below.

\subsection{Normalization}
\label{sec:normalization}

As explained in \secref{back:reconstruction}, the lowering of HighIR probe
operations to reconstruction code requires first normalizing the field definitions
that are being probed. 
We implement this process as a rewriting system on the bodies of \name{} operations
after fusion.

The basic strategy of normalization is as in the previous compiler: an \name{}
expression in normal form will have field reconstructions, possibly involving
derivatives of the kernels, at the leaves, probe
operations at the next level up, and then first-order tensor operations in the levels
above.
Space does not permit a complete presentation of the rules, but we highlight the
key rewrites here.
The full system is described in the first-author's Ph.D. dissertation~\cite{chiw:phd-thesis},
including a formal definition of the normal form and proofs that the
rewrite system terminates, preserves types, and produces terms in normal form.

Pushing probes below lifted operators is the purpose of the following rules. which distribute
the probes over unary and binary operations:
\begin{displaymath}
  \begin{array}{ccc}
    (-F_{\alpha})@x \Rightarrow -({F_\alpha @x}) & \quad &
    \sqrt{F_\alpha}@x \Rightarrow \sqrt{F_\alpha @x} \\
    \multicolumn{3}{c}{
      (e_1 \odot{} e_2)@x \Rightarrow (e_1@x) \odot{} (e_2@x) \quad
      \text{where $\odot\in\SET{{+}, {-}, {*}}$}
    }
  \end{array}
\end{displaymath}%
Similar rules existed in the previous compiler, but only for a very limited set
of operators.

We also need rules to apply differentiation to other operators and, in this way, push
it down toward the leaves.
These rules are justified by the semantics of tensor calculus~\cite{introTC}.
For example, we have multiple rewrite rules that embody the \emph{chain rule} from
calculus, such as
\begin{displaymath}
  \einpart{i}(\textbf{exp}(e))
  \Rightarrow
  \textbf{exp}(e) (\einpart{i} e)
\end{displaymath}%
and a rewrite for the quotient rule
\begin{displaymath}
  \einpart{i} \frac{e_1}{e_2}
  \Rightarrow
  \frac{(\einpart{i} e_1) e_2 -e_1 (\einpart{i}  e_2)}{e_2^2}
\end{displaymath}%
Other rules handle cases such as distribution over binary operators
\begin{displaymath}
  \einpart{i}(e_1 \odot e_2)
  \Rightarrow
  (\einpart{i} e_1) \odot (\einpart{i} e_2)
  \quad\text{for $\odot\in\SET{{+}, {-}}$}
\end{displaymath}%
and trigonometric identities
\begin{displaymath}
  \einpart{i} \arccos(e)
  \Rightarrow
  \frac{-(\einpart{i} e)}{\sqrt{(\lift{1}-(e*e)}}
\end{displaymath}%

Once the differenttiation operator is pushed down to an initial-field
definition, we rewrite the expression by moving the indices
on the differentiation operator to the reconstruction kernel.
\begin{displaymath}
  \einpart{\mu} (V_{\alpha} \circledast H^{\beta})
  \Rightarrow
  V_{\alpha} \circledast  H^{\beta \mu}
\end{displaymath}%

\subsection{HighIR optimizations}
\label{sec:high-opt}
In addition to normalization, we also include a number of domain-specific rewrites
the optimize terms in our HighIR rewriting system.
An important class of these are index-based reductions that allow us to eliminate
summations (\ie{}, effectively reduce the loop-nesting depth).
\begin{displaymath}
\sum_{jk}  \mathcal{E}_{ijk} \left(\einpart{ij} e\right)
  \Rightarrow
  \lift{0}
\end{displaymath}%
Two epsilons in an expression with a shared index can be rewritten
to deltas~\cite{math-for-physics}.
\begin{displaymath}
\sum_{i}  \mathcal{E}_{ijk} \mathcal{E}_{ilm}
  \Rightarrow
  \delta_{jl}\delta_{km} - \delta_{jm}\delta_{kl}
\end{displaymath}%
Application of a $\delta_{ij}$ to tensors and fields can be simplified.
\begin{displaymath}
  \begin{array}{ccccccc}
    \sum_j \delta_{ij} T_{j} \Rightarrow T_i & \quad &
    \sum_j \delta_{ij} F_{j} \Rightarrow T_i & \quad &
    \sum_j \einpart{j} \delta_{ij} e \Rightarrow \einpart{i} e
  \end{array}
\end{displaymath}%

Lastly, we also include rewrites that implement standard constant folding of arithmetic expressions.

One of the nice features of the rewriting system is that ``discovers'' identities without
having to have specific rules.
For example, the compiler will reduce the trace of the outer product of two vectors
to their dot product, and it will reduce $\nabla \bullet \nabla \times F$ to zero.
Another example is the simplification
\begin{displaymath}
  (a\times b) \bullet (c \times d) \Rightarrow (a \bullet c )(b \bullet d) -(a\bullet d)( b \bullet c)
\end{displaymath}%
(the right-hand side is preferred because it involves fewer arithmetic operations).
The \name{} operator created for the left-hand side has the body
\begin{displaymath}
  \sum_i (\sum_{jk}\mathcal{E}_{ijk} a_j b_k )(\sum_{lm} \mathcal{E}_{ilm} c_l d_m)
\end{displaymath}%
During normalization  the two epsilons with a shared index are reduced to deltas
\begin{displaymath}
  (\sum_{jklm} \delta_{jl} \delta_{km} a_j b_k c_l d_m)-\sum_{jklm}(\delta_{jm} \delta_{kl} a_j b_k c_l d_m)
\end{displaymath}%
The deltas reduction produces
\begin{displaymath}
  (\sum_{jk}a_j b_k c_j d_k)-\sum_{jk}(a_j b_k c_k d_j).
\end{displaymath}%
The summation binding method (discussed later in this section) produces
\begin{displaymath}
  (\sum_{j}a_jc_j)(\sum_k b_k d_k) - (\sum_j a_j d_j)(\sum_k b_k c_k)
\end{displaymath}%
which is the \name{} representation of $(a \bullet c)(b \bullet d) - (a \bullet d)(b \bullet c)$.

\subsection{Lowering to MidIR}
Once we have normalized the HighIR, we then need to lower the representation to MidIR.
Like HighIR, the MidIR also uses \name{} operators as a compact representation of loops
over tensor indices
After normalization, definitional field expressions are no longer necessary and can be eliminated.
Normalization also has the effect of removing the differential operator from the \name{}
expressions.
Thus the resulting MidIR representation involves computations on scalars and tensors, but not
fields.

\subsection{Lowering to LowIR}
In the Mid-IR phase of the compiler, operations between tensors, images, and kernels are
represented by \name{} operators.
The translation to LowIR expands the bodies of the \name{} operators
into vectorized code.
The inner iteration of an operator is usually mapped onto vector operations while any outer
iterations are unrolled into straight-line code.
Aggressive loop unrolling makes sense in our domain because the number of iterations is
typically two or three and the loop-nesting depth is usually one or two.

The LowIR assumes arbitrarily-wide vector operations; these are then mapped onto the
available hardware resources (\ie{}, 128, 256, or 512-bit wide vectors depending
on the target machine) as part of the translation to expression trees in the back end.

\subsection{Managing the size of the IR}
\label{sec:size-management}

The initial implementation of \name{} was found to support the existing
body of Diderot programs and benchmarks well with slight improvements
in efficiency~\cite{chiw:masters-thesis}.
There was an unexpected consequence, however from the switch to \name{}.
With \name{} in place, we were able to greatly increase the expressiveness
of the Diderot language (by adding more differentiation operators and
lifting tensor operations), which in turn enabled richer and more
sophisticated Diderot programs.
These new programs, which could never have been written in the previous
version of Diderot, stressed the compiler in unexpected ways that resulted
in lengthy compile times and failures to compile.

In hindsight, the cause of the problem was obvious.
The use of fusion on HighIR is necessary to group expressions so that we can effectively
apply normalization and lowering.
But as noted above, however, it has the effect of duplicating code.
In the High and Mid IRs, this code duplication is not a serious problem, but when we
lower MidIR to LowIR the duplicated terms get expanded and we get exponential code blowup in some cases.
To address this problem, we developed several transformations that we apply to the MidIR
immediately after lowering from HighIR.
The goal of these transforms is break apart larger \name{} operators into smaller ones,
where the SSA value-numbering optimization will then be able to eliminate many of them
as redundant.

\subsubsection{Split}
\label{sec:split}

The \optsplit{} transformation decomposes a large \name{} operator into smaller and simpler ones
--- in some respects, it is the inverse of fusion.
The basic idea of \optsplit{} is fairly simple, but the tricky part is correctly tracking
the indices so that we preserve the semantics of the expression.
We define the \emph{shape} of an \name{} expression $e$ to be the ordered list of index variables that
are free in $e$; these variables are bound in an outer context (either an outer summation or the
index map of the containing operator) and the order is given by a top-down, left-to-right ordering
of their binding sites~\cite{chiw:phd-thesis}.
For example, the shape of the expression $T_{j,i}$ in the context
$\EinOp{\mathit{xs}}{\cdot}{i\leq{}n_1,j\leq{}n_2}$
is $i\leq{}n_1,j\leq{}n_2$, while the shape of the expression
$\sum_{k\leq{}n_3} T_{i,k}$ in the same context is just $i\leq{}n_1$.

By splitting large \name{} operators into smaller ones, it provides opportunities
for the value-numbering pass to eliminate redundancy.
For example, the subexpression $e$ is embedded inside a larger \name{} operator that can be split apart 
\begin{displaymath}
  t_2 = \EinApp{A,B}{\left(\sum\limits_{\sigma_1} e\right) \odot e'}{\sigma_2}{a,b,\ldots}
  \Rightarrow
  \begin{array}{l}
    t_1 = \EinApp{B}{e}{\sigma'}{b} \\
    t_2 = \EinApp{A,T}{\left(\sum\limits_{\sigma_1} T_{\alpha}\right) \odot e'}{\sigma_2}{a, t_1}
  \end{array}
\end{displaymath}%
This transformation pulls the inner expression $e$ out to be its own operator  where $\sigma'$ is the shape of $e$.
In the original expression the expression $e$ is replaced by the tensor parameter $T_{\alpha}$ and $\alpha$ correspond to the index variables in $\sigma'$.
The parameter $B$ occurs in e and is moves to \name{} operator $t_1$, but $A$ does not and stays with \name{} operator $t_2$.

\subsubsection{Slice}
\label{sec:slice}
A sliced field is a field term that has at least one constant index ($c$) to indicate a field component.
Different sliced field terms are distinct (\ie{}, $F_{01} \neq{} F_{11}$),
but since they depend on the same source data they will generate many of the same base computations.
To avoid this redundancy, we decompose an \name{} operator with a sliced field term
into a structure that is more general, but still applies the necessary tensor slicing operator.
With this approach we can capture common terms before translating to lower-level constructs while still maintaining the mathematical meaning behind the original terms.
The decomposition is as follows:
\begin{displaymath}
  g = \EinApp{F,x}{\einpart{\beta} F_{c\alpha}(x)}{\hat{\alpha}\hat{\beta}}{f,x}
  \Rightarrow
  \begin{array}{l}
    t = \EinApp{F,x}{\einpart{\beta} F_{\gamma}(x)}{\hat{\gamma}\hat{\beta}}{f,x} \\
    g = \EinApp{T}{T_{c\alpha\beta}}{\hat{\alpha}\hat{\beta}}{t}
  \end{array}
\end{displaymath}%
where $\hat{\gamma} = i\leq{}n, \hat{\alpha}$, the field variable $F$ has type  \lstinline[mathescape=true]!field#k(d)[$n :: \varsigma$]!, and the tensor $x$ has type
\lstinline[mathescape=true]!tensor[d]!. 
The result is a pair of \name{} operators: an unsliced field probe and a
tensor-valued \name{} operator.
The defintion of $g$ captures the components of the probe that are of interest.
Like, the split method, the slice method creates a representation that will produce a smaller number of lower level constructs.

\subsubsection{Summation Binding}
\label{opt:shift}
Each summation operator represents a loop nest that will be unrolled when lowered to LowIR.
Thus moving operations outside the summation can avoid subsequent code duplication.
This transformation is essentially loop-invariant code hoisting for the special case
of summations.
We identify loop invariants by looking at the indices.
For example, consider the term $\sum\limits_{\sigma}(T_{\alpha}*e)$ if the property
$\forall i\in \alpha. i \not \in  \sigma $ holds then we know that $T_{\alpha}$ is invariant
and can be lifted.
\begin{displaymath}
  \sum\limits_{\sigma}(T_{\alpha} * e)
  \Rightarrow
  T_{\alpha} * \sum\limits_{\sigma}(e)
\end{displaymath}%
When there are nested summations, then the method applies additional analysis
to see if the summation can be converted into the product of independent
summations as demonstrated in the example in \secref{sec:high-opt}.

\subsubsection{Measuring Size Reduction}
\begin{figure}[t]
  \begin{center}
    \includegraphics[width=4.2in]{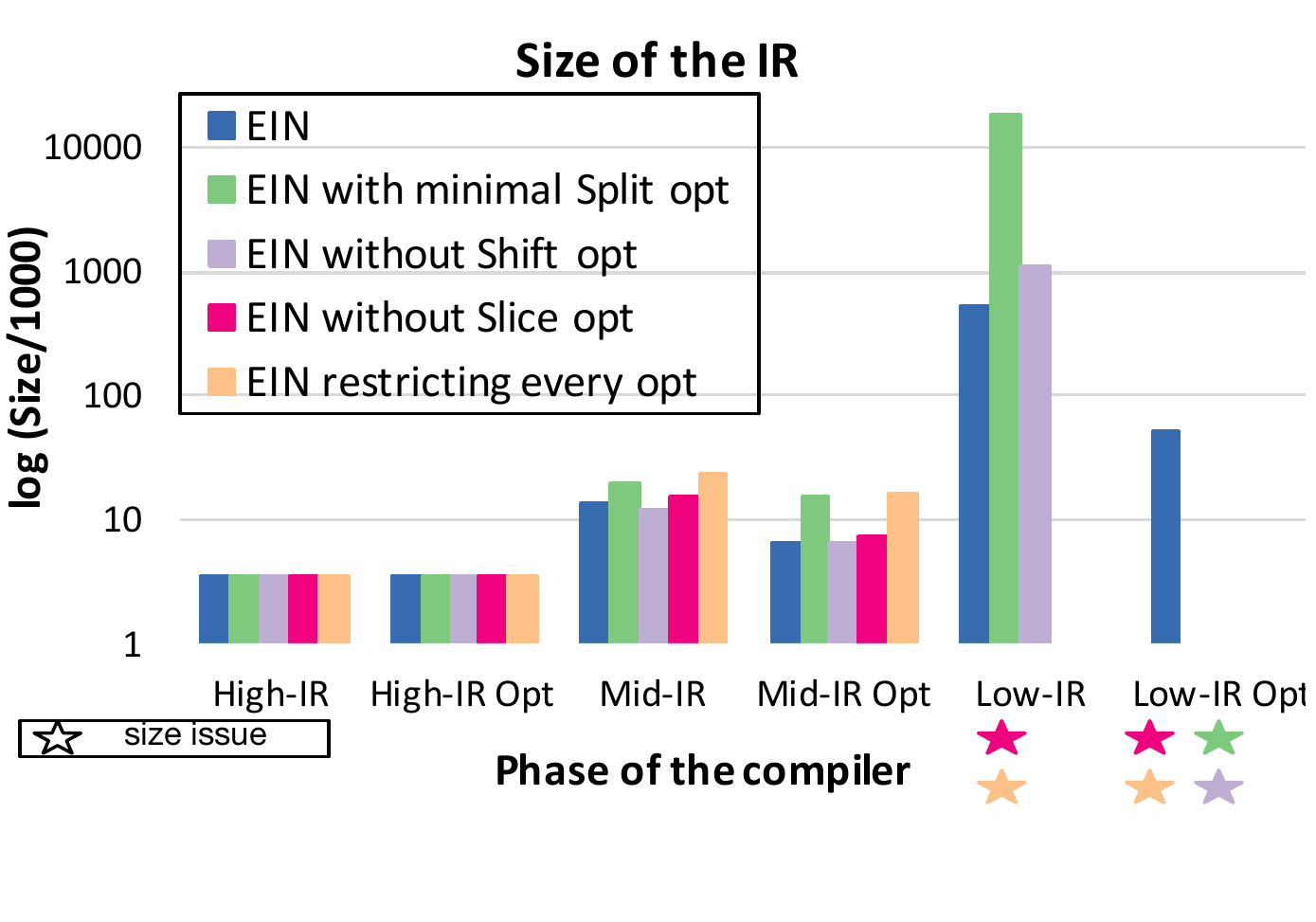}
  \end{center}%
  \caption[Size of programs at different phases of the compiler]{
    The graphs shows the size of the intermediate representation at different
    points in the compilation of the the rsvr program.
    The size is measured by counting the number of expression-tree nodes
    used to represent the program.
    ``EIN'' is the baseline with full optimizations.
  }
  \label{fig:ir-sizes}
\end{figure}

The techniques described in this section were motivated by a need to manage
the size of the IR during compilation.
To evaluate the effectiveness of the techniques, we instrumented the compiler
with code to measure the size of the IR at six different points in the
compilation pipeline.
\figref{fig:ir-sizes} plots the IR sizes for the program ``rsvr,'' which is
a Diderot program that computes ridge surfaces using higher-order programming.
The first thing to note is that the program only compiles if all the optimizations
are enabled.
The other thing to note is that the size problems are localized to the LowIR
following lowering from MidIR.
Measurements of other higher-order programs produce similar results.
Thus our strategy of applying transforms to the MidIR and then applying value
numbering reduces the growth in the IR to a manageable level.

%
\section{Benchmarks}
\label{sec:benchmarks}

While it is clear that the \name{} supports a more expressive programming model
than the previous compiler, there is still an open question of what effect it
has on both compile times and run times.
We are also interested in the measured effect of the various transformations
described in \secref{sec:size-management}.
To evaluate the impact and cost of the compilation techniques we measure the
compilation and execution time for various benchmarks.
For the benchmarks we use a mix of Diderot programs with varying degrees of
tensor math.
We present the results of three experiments:
the first compares the performance of \name{} with the previous compiler;
the second measures the impact of different compiler settings; and
the third compares higher-order programs with their first-order equivalents.

\subsection{Experimental Framework}
The benchmarks were run on an Apple iMac with a 2.7 GHz Intel core i5 processor,
8GB memory, and OS X Yosemite (10.10.5) operating system.
Each benchmark was run 10 times and we report the average time in seconds.

The benchmarks are presented in the figures in order of increasing mathematical
complexity.
Some of the visualization concepts that inspired the benchmarks are described
in \secref{sec:applications}.
Benchmarks ``illust-vr,'' ``lic2d,'' ``Mandelbrot,''``ridge3d,'' and ``vr-lite-cam''
are small examples that were used to evaluate the previous compiler~\cite{diderot:pldi12}.
The benchmarks ``mode,'' ``canny,'' and ``moe'' are more sophisticated and were
described in a previous paper~\cite{diderot:vis15}.
``Mode'' finds lines of degeneracy in a stress tensor
field revealed by volume rendering isosurface of tensor mode;
``Canny-edges'' computes Canny Edges;
and ``Moe'' volume renders
isocontours found using Canny Edges~\cite{canny1986}.

The benchmarks ``dec-crest,'' ``dec-grad,'' ``rsvr,'' and ``mode-rig'' have not
been featured in previous work because they were not expressible in previous
versions of the compiler.
The programs ``dec-crest'' and ``dec-grad'' are approximations to illustrate the crest
lines on a dodecahedron.
Programs ``mode-rig'' and ``rsvr'' are both programs created to  measure ridge lines.
The micro-benchmarks ``det-grad,''``det-hess,''  and ``det-trig'' compute a single
property: the gradient, hessian, and various functions computed on the determinant of a field.
Unlike the other benchmarks, these programs are not visualization programs and have
negligible runtimes, which we omit.

\subsection{Comparing compilers}
\label{bench:ein}

\begin{figure}[p]
  \begin{center}
    \includegraphics[width=4.4in]{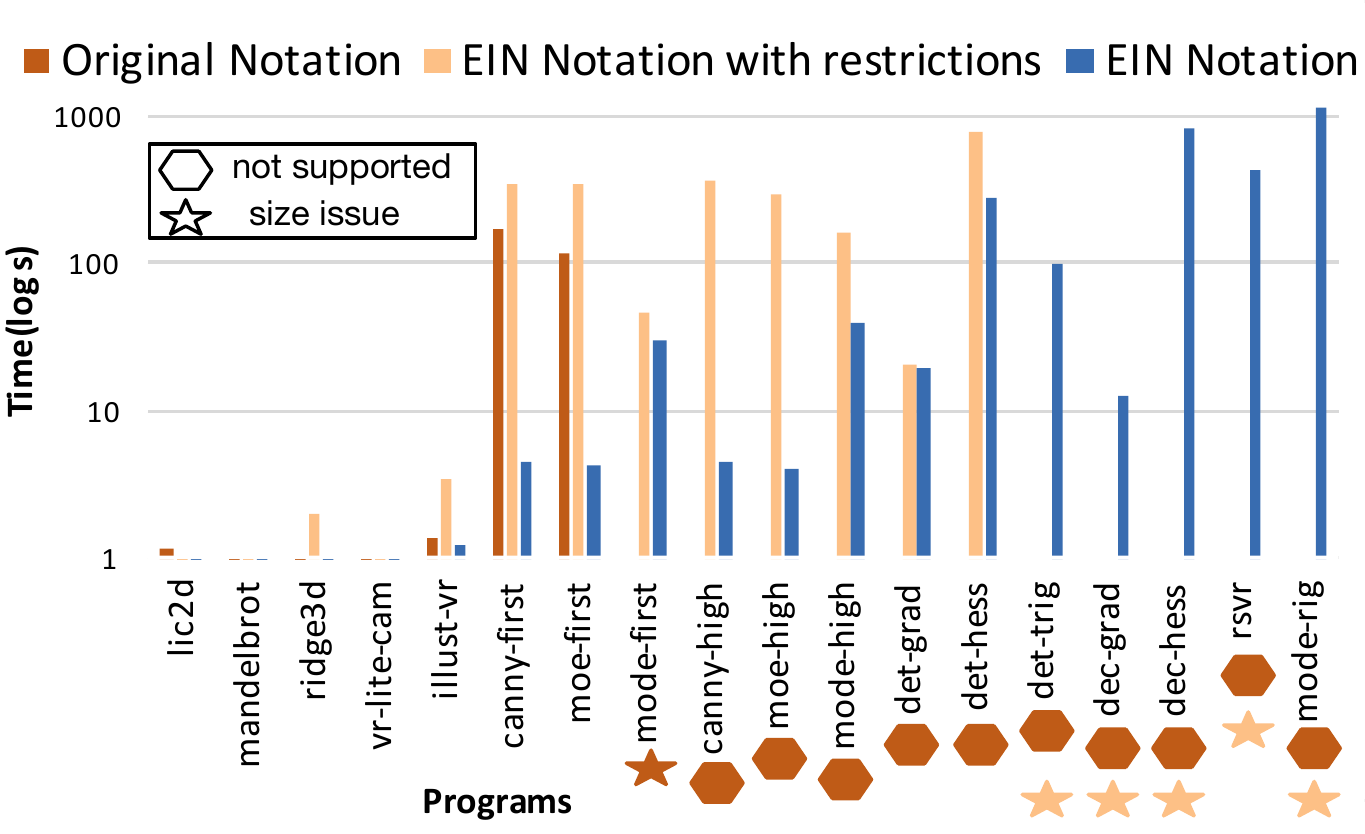}\\
    \textbf{Compile times}\\[2em]
    \includegraphics[width=4.4in]{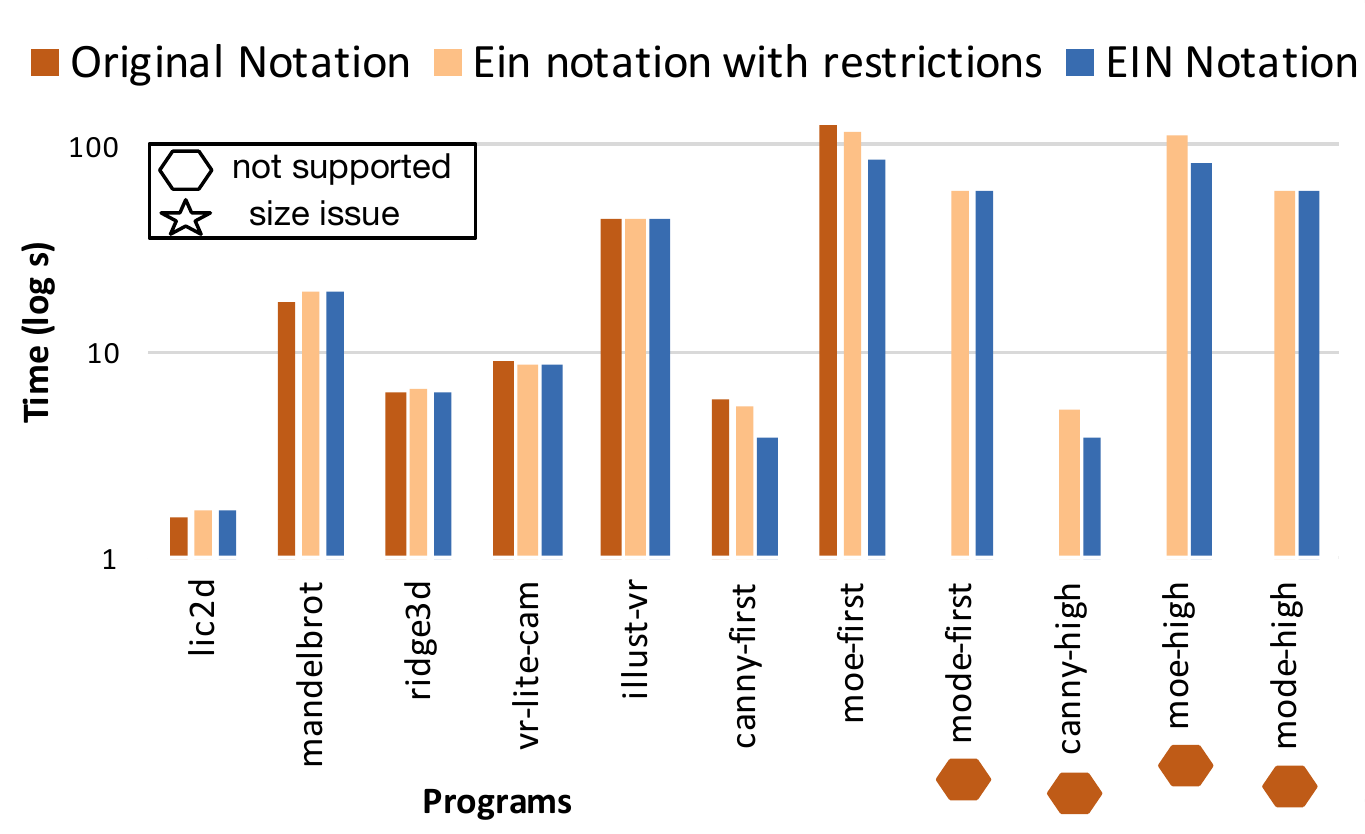}\\
    \textbf{Execution Times}
  \end{center}%
  \caption{
    Comparing three different compilers.
    The ``Original''version of the compiler does not use the \name{} IR.
   ``\name{}  with restrictions'' is the more naive implementation of \name.
   ``\name{}'' is the baseline with the \name{} IR with full optimizations applied.
   Fully implementing \name{} allows more programs to compile than previously possible.
  }
  \label{fig:naive}
\end{figure}

Our experiments consist of eighteen benchmarks run on three versions of the compiler.
The first is the original direct-style compiler~\cite{diderot:pldi12}.
The second and third include the design and implementation techniques included in this
paper, but the second version imposes restrictions that are meant to reflect the first
\naive{} implementation of \name{}.
We measure the time it takes for the programs to compile.
For the programs that compile on at least two versions, we also compared the run times.
The results are given in \figref{fig:naive}.
We use a hexagon to indicate that the compiler did not support the features used by a program
and we use a star symbol to indicate a program that did not compile; the
symbols are color coded by compiler.

Not surprisingly, the original compiler was not able to compile many of the programs
because of expressiveness issues.
And for those that it could compile, its compile times were often significantly slower
than \name{}.
The restricted form of \name{} was uniformly worse than \name{} in compile times
and could not compile the most complex examples.
It should be noted, however, that some of the more advanced programs took close to
20 minutes to compile in \name{}, so there is room for improvement.
The execution-time story is more balanced.
The original compiler produces slightly faster code for some benchmarks, but the \name{}
compiler produced the fastest code for the more complex programs.
These results suggest that we have not lost performance by switching to the more expressive
IR.

\subsection{The Effect of compiler settings}
\label{bench:effect}

\begin{figure}[p]
  \begin{center}
    \includegraphics[width=4.4in]{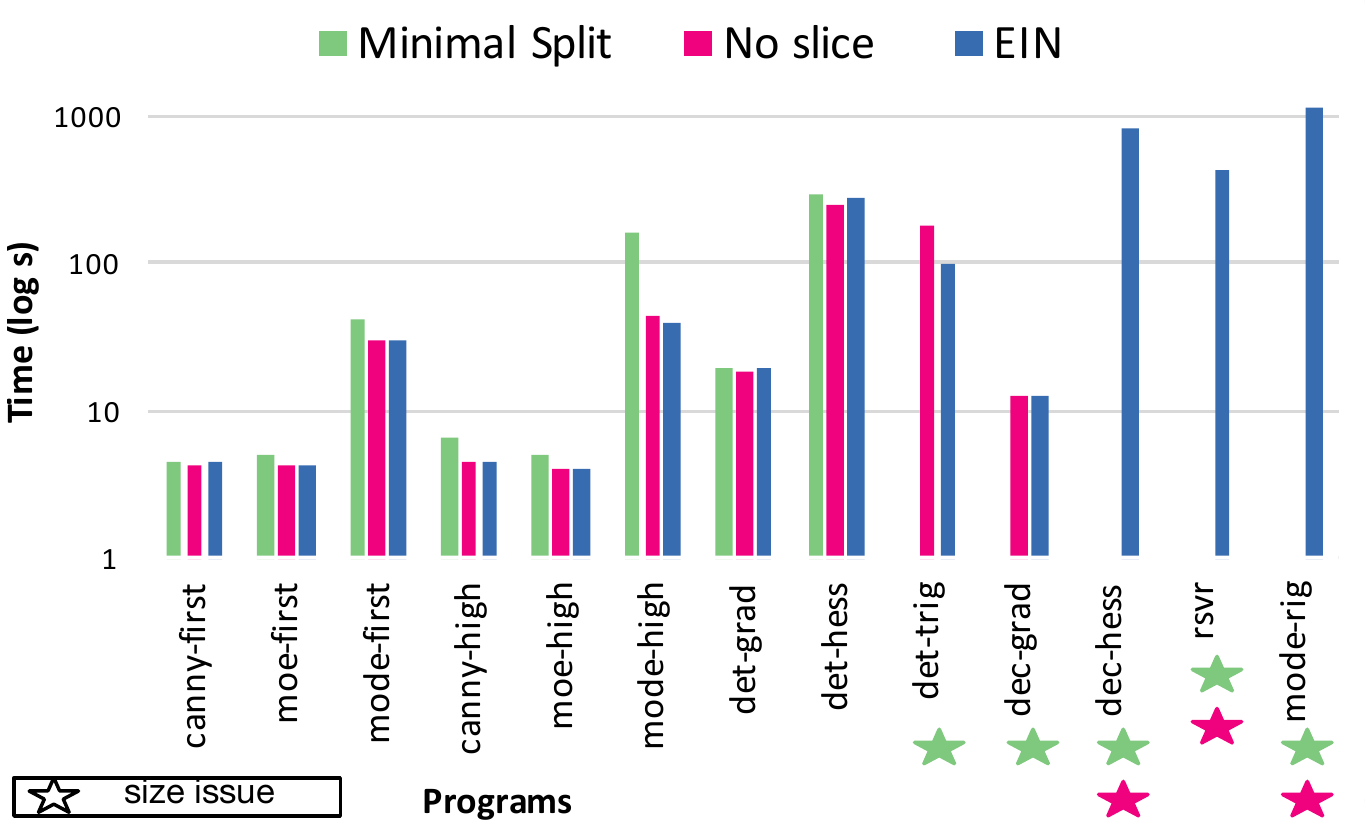}\\
    \textbf{Compile times}\\[2em]
    \includegraphics[width=4.4in]{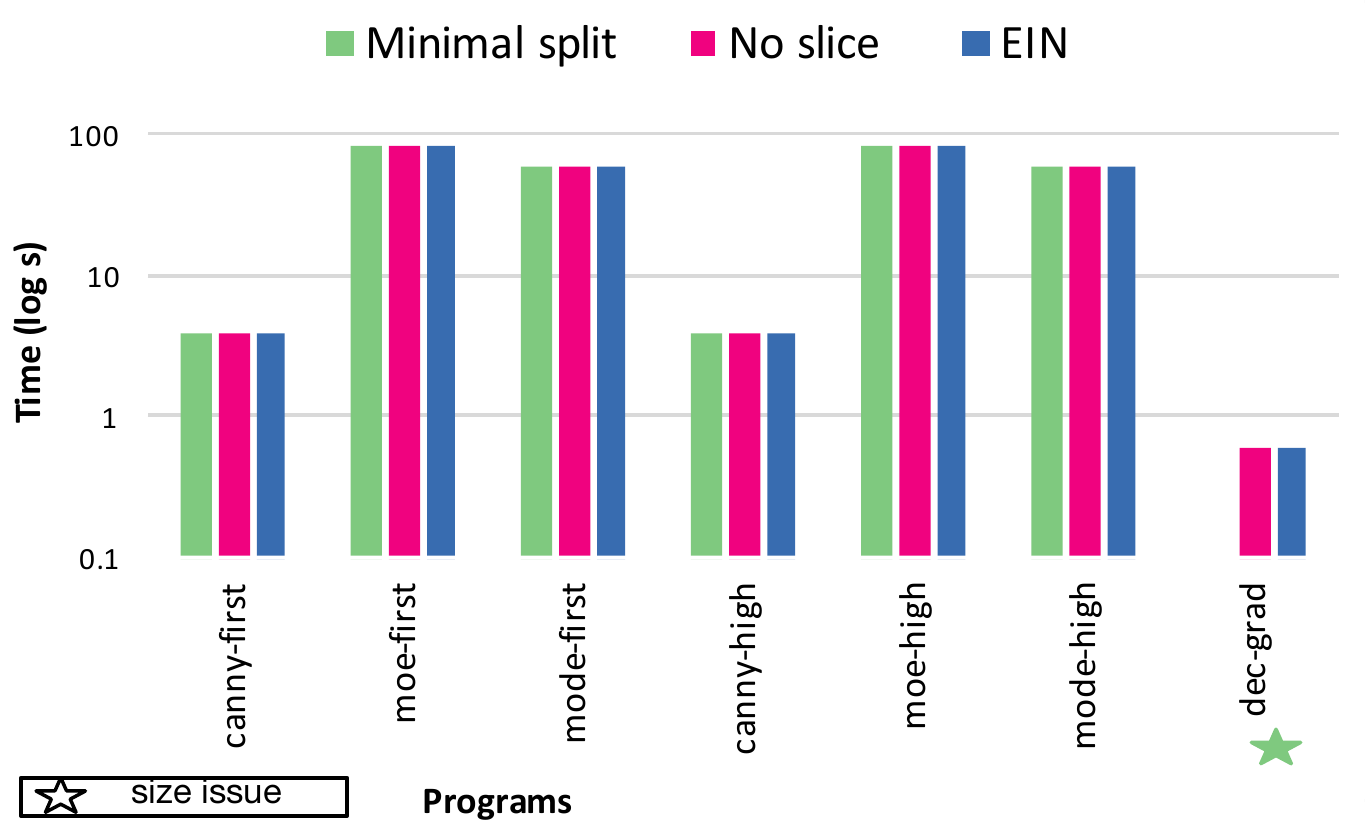}\\
    \textbf{Execution Times}
  \end{center}%
  \caption[Measurement of applying optimizations techniques]{
    Compile and run time measurements when implementing \optslice{}
    and \optsplit{} on High-IR.
    Doing no amount of splitting prevents most of these programs from
    compiling so instead we measure its impact  by limiting it, ``Minimal \optsplit{}''.
    \name{} is the baseline with both \optsplit{} and \optslice{} enabled.
  }
  \label{fig:slice}
\end{figure}%

As we have discussed previously, a \naive{} application of our transformations causes
unacceptable space blowup.
To address the space problem we developed techniques to reduce the size of the IR resulting
from lowering passes.
While their implementation might allow more complicated Diderot programs to be compiled,
we want to evaluate their cost and benefit for programs that could already compile.
In the following, we evaluate the effectiveness of these techniques together and isolated
at different steps in the compiler.

\paragraph{Application to higher-order constructs}
Techniques \optsplit{} and \optslice{} are effective at reducing the size of the program
by finding common subexpressions or reducing field terms.
\figref{fig:slice} measures the effectiveness of applying \optsplit{} and \optslice{}
on a HighIR \name{} operator.
The \optslice{} technique is necessary to compile three of the thirteen benchmarks.
\optsplit{}  is the most consequential technique.
Limiting the application of the method stops five of the programs from being able to compile.
 Techniques Split and Slice do not add take a considerably longer time to run.

\begin{figure}[p]
  \begin{center}
    \includegraphics[width=4.4in]{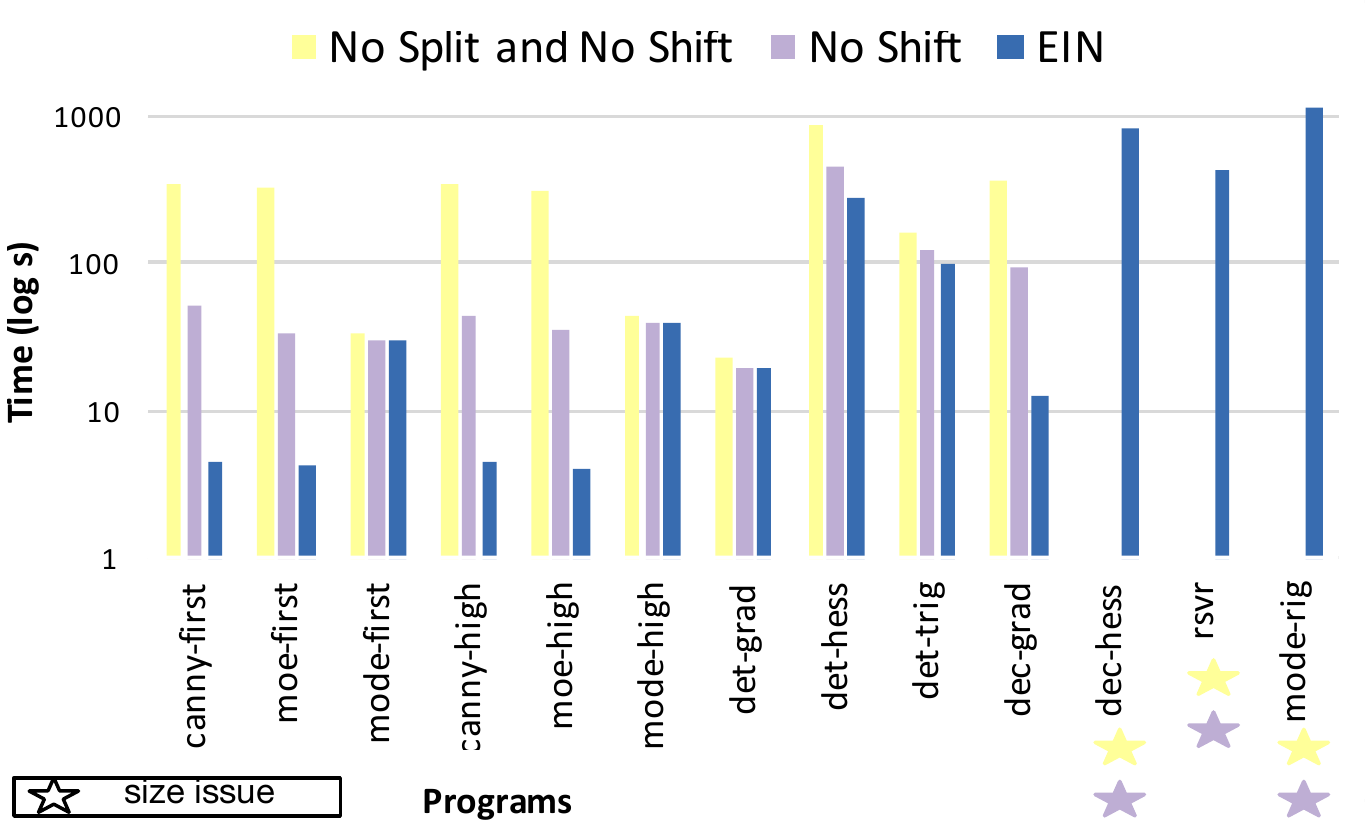} \\
    \textbf{Compile times}\\[2em]
    \includegraphics[width=4.4in]{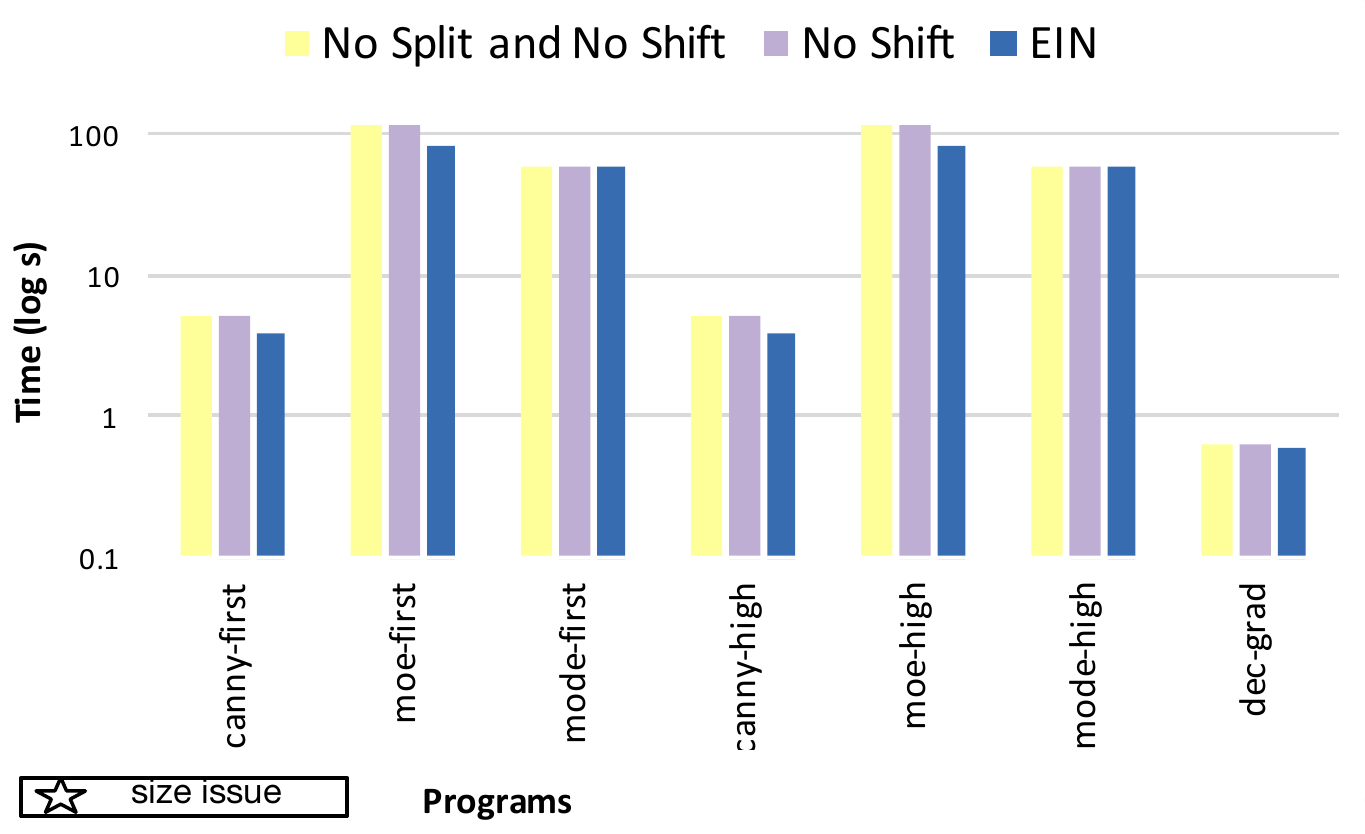} \\
    \textbf{Execution Times}
  \caption{
    Compile and run time measurements for implementing  \optshift{} and \optsplit{}
    techniques on reconstructed field terms. \label{tbl:fieldexp} \name{} baseline
    includes the application of both \optshift{} and \optsplit{}.
  }
  \end{center}
  \label{fig:shift}
\end{figure}%

\paragraph{Application to lower-order constructs}
\label{sec:benchlow}
 We measured the effect of optimizations at a later phase of the compiler.
%
The implementation of the techniques directly affect compile time.
Applying optimizations  \optshift{} and \optsplit{} together  offers a  consistent
speed-up on the execution time and compile time for all thirteen benchmarks.
Five programs experienced at least a 20-times improvement in compile time speed-up
and four of the benchmarks offered at least a 1.3 speed up on execution time while
the rest saw no change in the execution time.

\subsection{Higher-order versus first-order programs}
\begin{figure}[p]
  \begin{center}
    \includegraphics[width=4.4in]{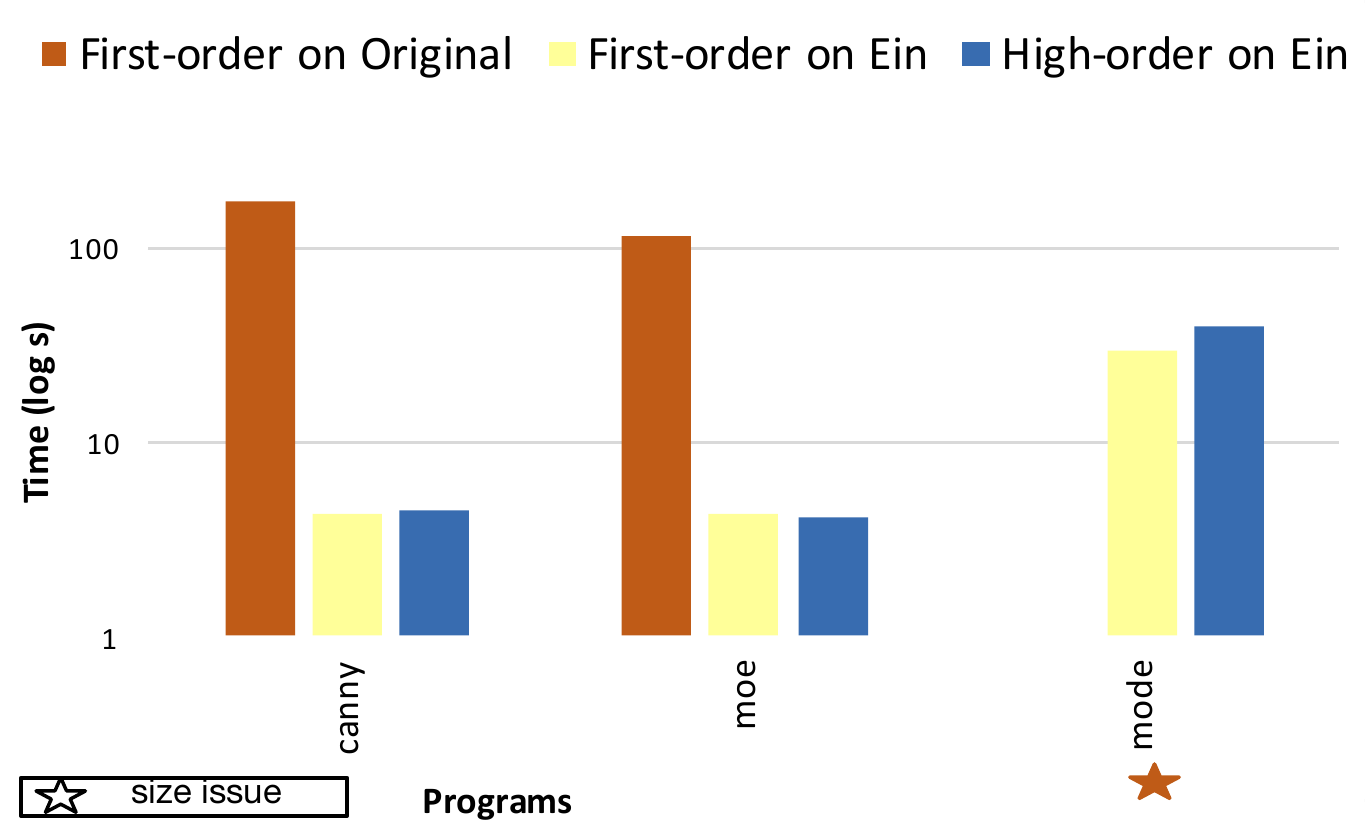} \\
    \textbf{Compile times}\\[2em]
    \includegraphics[width=4.4in]{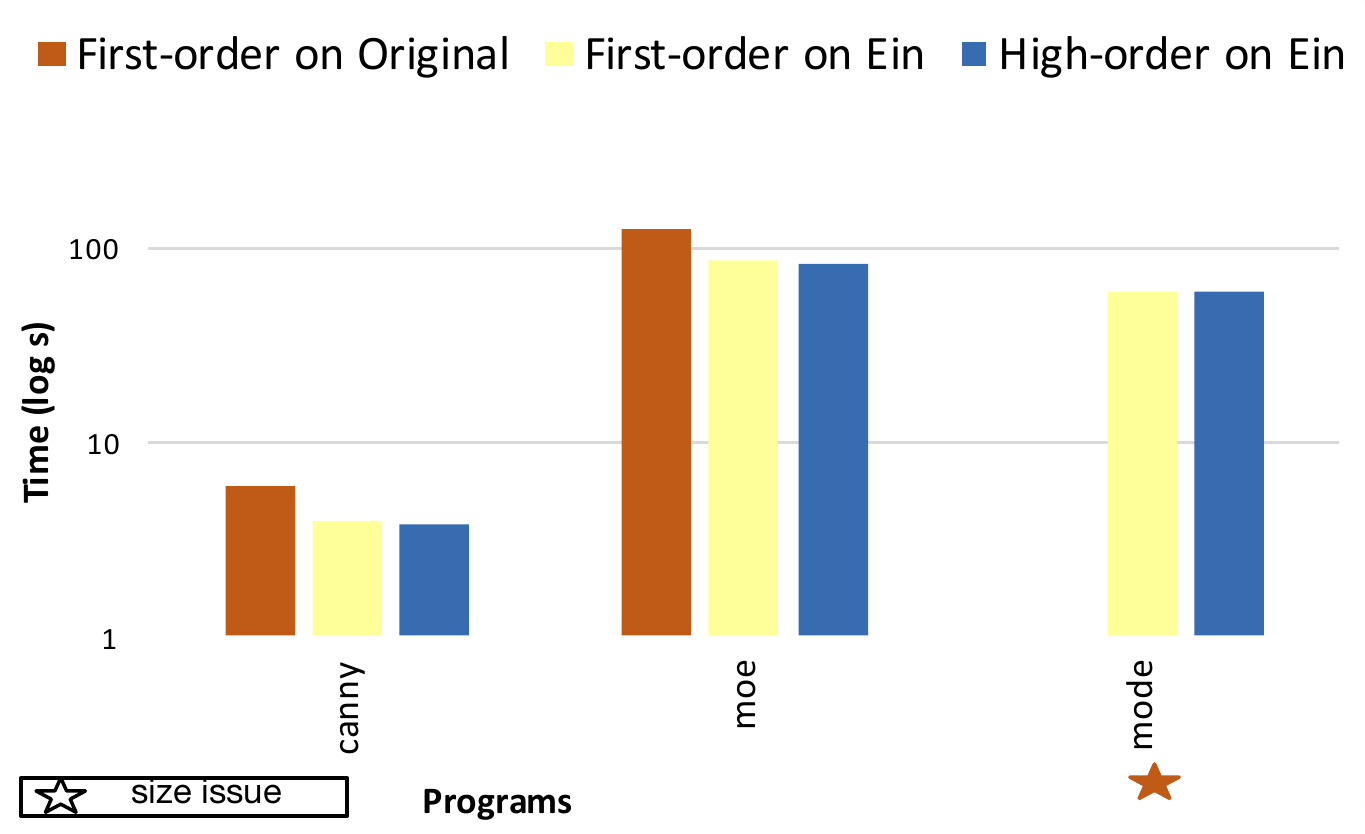} \\
    \textbf{Execution Times}
  \end{center}
  \caption[Programs written with first or higher order  operators]{
    A comparison of hand-derived first-order programs with their
    higher-order equivalent.
    The original compiler can only compile the first-order versions,
    while the compiler with \name{} can compile both versions.
  }
  \label{fig:firstorder}
\end{figure}

For this experiment we wrote both a higher-order and first-order version
of three programs.
The higher-order programs use lifted tensor operators on fields, whereas the
first-order versions only applies tensor operators to tensors.
\tblref{tbl:lines} gives a breakdown of the relative program sizes in lines of code
and \figref{fig:firstorder} shows the compilation and execution times for
the two versions.
For the first-order version, we measured the performance with both the previous
compiler and with \name{}.

The results demonstrate that using higher-order operators in Diderot does not
impose a performance penalty in either compile time or execution time.
(the one exception was the first-order ``mode'' program, which compiled slightly
faster than the higher-order program).

\begin{table}[t]
  \caption{
    The table reports the lines of code written in a program.
    Two version of the program are written; ``FO'' first-order and `HO'' high-order.
    The full length of the program is reported as well as the core parts of the
    program.
    The table also measures the number of lines used in the field definitions,
    function declarations, and the central update method.
  }
  \label{tbl:lines}
  \begin{center}
     \begin{tabular}{l||cc|cc|cc|cc|}
       & \multicolumn{2}{c|}{Program length}
       & \multicolumn{2}{c|}{Field Definitions}
       & \multicolumn{2}{c|}{Functions}
       & \multicolumn{2}{c|}{Update Method} \\
      \hline
      Program name & FO & HO & FO & HO & FO & HO & FO & HO \\
      \hline
      Mode
      & 125 & 75
      &  3 & 4
      & 19 &3
      & 61&25 \\
      Moe
      & 84 & 76
      & 5 & 4
      & 4  & 3
      & 29& 23\\
      Canny
      & 77 & 77
      &2  &5
      & 2 & 2
      & 26& 23
    \end{tabular}
  \end{center}
\end{table}%


%
\section{Applications}
\label{sec:applications}

Scientists use image analysis to extract features that allow better understanding
of their image data.
These features can be based on individual pixels, detection of regions with specific
shape, time, and transformations of the data~\cite{medical-img-processing}.
Our work in the \name{} IR makes it easier to develop analysis algorithms that exploit
higher-order operations.
This section describes two visualization techniques~\cite{diderot:vis15,ein:cpc-paper}
that illustrate the effect of switching to the \name{} IR in the compiler.


Vector fields arise in the analysis of fluid flow; properties of the derivatives of the
vector field characterize important features (such as vortices) in the flow.
For example, the curl $\nabla \times V$, indicates the axis direction and magnitude of
local rotation.
One definition of vortices
identifies them with places where the flow direction $\frac{\mathbf{V}}{|\mathbf{V}|}$
aligns with the curl direction
$\frac{\nabla\times\mathbf{V}}{|\nabla\times\mathbf{V}|}$~\cite{helicity}.
\emph{Normalized helicity} measures the angle between these directions:
\begin{lstlisting}[mathescape=true]
    field#3(3)[]H = (V/|V|) $\bullet$ ($\nabla\times$V/|$\nabla\times$V|);
\end{lstlisting}
in terms of the vector field \lstinline!V!.
The \name{} IR allows the core computation to be expressed directly as operations on fields.

Material properties, such as diffusivity and conductivity, vary locally in
magnitude, orientation, and directional sensitivity, so they are modeled with second-order
tensor fields.
Visualizing the structure of tensor fields typically depends on measuring
various tensor invariants, such as
\emph{anisotropy}: the magnitude of directional dependence.
For example, neuroscientists study the architecture of human brain white matter with
diffusion tensor fields computed from MRI~\cite{basser1996}.
A popular measure of diffusion anisotropy,
``fractional anisotropy'' can be directly expressed in Diderot as:
\begin{lstlisting}[mathescape=true]
  field#4(3)[3,3]E = T - trace(T)*identity[T]/3;
  field#4(3)[]A = sqrt(3.0/2.0)*|E|/|T|
\end{lstlisting}%
This code measures the magnitude of the purely anisotropic
\emph{deviatoric tensor} relative to the tensor field \lstinline!T! itself.

Subsequent visualization or analysis will
typically require differentiation, such as the first derivatives
needed for shading renderings of isocontours, or the
second derivatives needed for extracting ridge and valley
features.
Generating expressions for 
$\nabla$ \lstinline!H!, and $\nabla$ \lstinline!A! by
hand is cumbersome and error-prone, whereas \name{} allows Diderot to easily handle these
and more.

%
\section{Related Work}
\label{sec:related}

\subsection{Einstein Index Notation}

\name{} is inspired by\emph{Einstein Index Notation}, which is a concise  written
notation for tensor calculus invented by Albert Einstein~\cite{Einstein:general-theory}.
Einstein Index Notation uses repeated indices to represent implicit summations
(thus it is sometimes called the summation convention).
It has be used to represent a wide array of physical quantities
and algorithms in scientific
computing~\cite{einstein-for-arrays,einstein-notation,math-for-physics,introTC,brief-tensor-analysis,tensor-analysis}.

Various designers have studied the ambiguities and limitations of the notation to extend
its uses on paper and to develop a grammar and semantics for implementation.
A  part of the ambiguity in index notation is related to the implicit summation.
Therefore, various notational definitions are created to suppress
summation~\cite{einstein-for-arrays}.
These include using notation to differentiate between types of indices, using a
no-sum operator~\cite{einstein-notation}, and
differentiate between indices that repeat exactly twice~\cite{introTC,tensor-analysis}.
\name{} notation is a compiler IR so the goal is for the IR to supply enough
information to represent the wide range of operations.
\name{} notation uses an explicit summation symbol so it can
explicitly set boundaries for diverse operations.
The notation leads  to more book-keeping but allows more expressivity.

There is existing work on implementing index notation in a surface language.
Ahlander describes supporting index notation for C++~\cite{einstein-for-arrays} and
Egi supports adding index notation to a functional programming
language~\cite{scheme-tensor-notation}.

Ongoing work to implement EIN on the surface level of Diderot is described
in the first author's dissertation~\cite{chiw:phd-thesis}.
Being able to support an index-like notation directly in Diderot could be beneficial
to quick prototyping and debugging.
Writing directly in index notation allows the developer to specify computations that
may be difficult to replicate with existing surface level operators.
It could also help a developer create intricate test cases to more rigorously test the
compiler implementation.

\subsection{Intermediate representations and optimizations}

Domain-specific languages can offer several benefits; the syntax and type system can be
designed to meet the practice and expectation of domain experts; the compiler can
leverage common domain-specific traits; and the programming model can abstract away
from hardware and operating system features.
By using a domain specific language, the end-user can  write code that is
familiar (to them) and let the system focus on generating high-performance code.
This notion of developing a high-level mathematical programming model is also
emphasized in Diderot.
There are various domain-specific languages that provide a link between
mathematical algorithms and programming.
This section focuses on five domain-specific compilers that are more closely
related to our work (Spiral~\cite{spiral},  TCE~\cite{TCEOutofCore},
{TSFC}~\cite{TSFC-form-compiler}, COFFEE~\cite{crossLoop}, and UFL~\cite{UFL}).

The tensor contraction engine (\fontrelated{TCE}) created by Hartono, Albert \etal{}
supports a high-level Mathematica style language for the quantum chemistry
domain~\cite{Tiling,TCEOpt,TCEOutofCore}.
The class of computations are multi-dimensional summations over products of several arrays.
The computations have a large number of nested loops and an explosively
large parameter search space.
The calculations can require a larger space than available physical memory.
To address this issue, Hartono \etal{} developed algebraic transformations
to reduce operation counts.
Like TCE, the Diderot language supports summations of products over several arrays.
Unlike TCE, Diderot supports a high level of expressiveness  between
tensors and tensor fields and field differentiation.

\fontrelated{Spiral} is a DSL created for digital signal processing~\cite{operator-language,spiralScala,spiral}.
Its design encapsulates significant mathematical knowledge of algorithms used
in digital signal processing.
P{\"{u}}schell \etal{}  addresses the goal of doing the right transformation at
the right level of abstraction.
Its implementation uses three levels of IR: ${SPL}$ to represent the signal processing
language, $\Sigma {SPL}$ to express loops and index mapping, and \fontrelated{C-IR}
for code level optimization.
$\Sigma SPL$  does loop merging and create complicated terms that are simplified
with a set of rewriting rules~\cite{SigmaSPL}.
\name{} and  $\Sigma \text{SPL}$ both use summation expressions to represent loops,
but the Spiral IR expresses  algorithms, while the \name{} IR expresses general
tensor calculus.
As with Spiral, Diderot allows its users to focus more on the mathematics and
allows the system to generate high-performance code for their platform.
Although Spiral provides a powerful mathematical model, it is targeted at a
somewhat different domain of signal processing.

The Unified Form Language (\fontrelated{UFL}) is a domain-specific language for
representing weak formulations of partial differential equations~\cite{UFL}.
UFL is  most closely similar to \name{}.
At its core  both UFL and \name{} support tensor algebra, high-level expressions
with domain-driven abstraction, and offer differentiation (automated vs. symbolic).
The projects address different domains, UFL is a language for expressing variational
statements of PDEs and Diderot is a language for scientific visualization and
image analysis.
UFL creates an abstract representation that is used by several form compilers to
generate low-level code.
Therefore, UFL avoids optimizations that a form compiler might want to perform and
instead sticks to a set of ``safe and local" simplifications.
Diderot controls the entire pipeline from surface language to code generation and
so it does have the opportunity to do optimal rewriting at a higher level.

\fontrelated{TSFC}~\cite{TSFC-form-compiler}  is a form compiler  that takes input from UFL.
The TSFC compiler uses an intermediate representation, called GEM.
Like EIN, GEM has a notation to represent index summation and tensor components.
GEM does not represent the rich range of tensor and field operators, that EIN does.
Part of the reason, is because  the symbolic differentiation and necessary rewriting
is done by UFL and as input TSFC sees a DAG.
Diderot maintains the tensor structure for as long as possible.
Diderot maps surface level operators directly into EIN notation and in that IR applies
rewrites and optimizations.

\fontrelated{COFFEE} is a domain-specific compiler for local assembly kernels\cite{crossLoop}.
The computation made by COFFEE is key to finding numerical solutions to partial
differential equations.
COFFEE computes the contribution of a single cell in a discretized domain for a linear
system to approximate a PDE.
The entire computation is discretized into a larger number of cells, making the time to compute this computation important.
To enable better performance COFFEE applies  optimizations.
COFFEE applies optimizations 
on a scalarized tensor expression tree.
Like COFFEE, \name{} does loop-invariant code motion and expression splitting on tensor expressions.
Unlike COFFEE, \name{} applies optimizations at a high level
to exploit the mathematical properties of the computations on higher-order tensors before flattening

There is extensive work on languages, libraries, and tools focused on efficient execution
of math operators.
Basic Linear Algebra Subprograms (BLAS)~\cite{level-3-blas} and Linear algebra PACKage (LAPACK)\cite{Anderson:1990:LPL:110382.110385} are popular high performance linear algebra libraries.
Build to order (BTO) specializes on matrix algebra~\cite{Nelson:2015:RGH:2786970.2629698}.
FLAME implements linear algebra algorithms and methods in the sequential world~\cite{Gunnels:2001:FFL:504210.504213}.
FGen focused on high performance convolution operators or finite-impulse-response (FIR)
filters~\cite{Stojanov:2014:AVA:2627373.2627376}.
LGen is a compiler to translate small-scale  basic linear algebra
expressions to C functions~\cite{linear-algebra-compiler}.
Like the previous work, Diderot generates code for first-order math operators on tensors.
Unlike the previous work, Diderot focuses on higher-order operators, such as
differentiation on continuous tensor fields.
Also, the tensors in our domain are small vectors and matrices.
In the future, our goal is to better leverage the tensor structure to generate faster code.


%
\section{Conclusion}
\label{sec:concl}

In this paper we have described a new IR that is used in the Diderot compiler.
This IR allows us to support a much richer level of mathematical abstraction
than was previously possible.
The key idea is to use a hybrid approach to the IR design; combining the advantages
of a normalized representation (SSA) with a compact representation of loop nests
based on expression trees.
We have motivated and described our approach and have presented performance
results as well as given examples of its utility.

While the techniques that we describe are implemented in the Diderot compiler,
we believe that they could be useful for implementing other DSLs that support
high-level mathematical programming models.

\section*{Acknowledgments}
Portions of this research were supported by National Science Foundation
awards CCF-1446412 and CCF-1564298.
The views and conclusions contained herein are those of the authors and should
not be interpreted as necessarily representing the official policies or
endorsements, either expressed or implied, of these organizations or the
U.S.\ Government.

\newpage
\bibliographystyle{splncs03}
\bibliography{gstrings-short,gdiderot,refs}

\end{document}